\documentclass[prx,twocolumn,superscriptaddress,citeautoscript,showpacs,amsart]{revtex4-2}

\usepackage{graphicx}
\usepackage{multirow}
\usepackage{physics}
\usepackage{color}
\usepackage{bm}
\usepackage{times}
\usepackage{amsmath,bm,amsfonts}
\usepackage{dcolumn}
\usepackage{graphicx}
\usepackage{latexsym}
\usepackage{cancel}
\usepackage{hyperref}

\usepackage{ulem} 
\usepackage{braket}

\def\ket#1{|#1\rangle }
\def\bra#1{\langle #1 |}
\def\n{\nonumber \\ }

\newcommand{\ma}{\sigma}
\newcommand{\V}{\vec}
\newcommand{\dt}{\delta}
\newcommand{\ep}{\epsilon}

\newcommand{\f}{\frac}
\newcommand{\ta}{\theta}

\newcommand{\dg}{\dagger}

\newcommand{\A}{\alpha}
\newcommand{\B}{\beta}

\newcommand{\w}[1]{\omega_{#1}}
\newcommand{\al}[1]{\langle #1 \rangle}

\newcommand{\alg}[1]{\begin{align}#1\end{align}}

\newcommand{\nm}{\nonumber \\&}
\newcommand{\nq}{\nonumber \\=&}
\newcommand{\p}{\partial}

\begin{document}

\title{Phonon thermal Hall effect in Mott insulators via skew-scattering by the scalar spin chirality}

\author{Taekoo \surname{Oh}}
\email{taekoo.oh@riken.jp}
\affiliation{RIKEN Center for Emergent Matter Science (CEMS), Wako, Saitama 351-0198, Japan}

\author{Naoto \surname{Nagaosa}}
\email{nagaosa@riken.jp}
\affiliation{RIKEN Center for Emergent Matter Science (CEMS), Wako, Saitama 351-0198, Japan}
\affiliation{Fundamental Quantum Science Program, TRIP Headquarters, RIKEN, Wako, Saitama, 351-0198, Japan}

\date{\today}

\begin{abstract}
Thermal transport is a crucial probe for studying excitations in insulators. 
In Mott insulators, the primary candidates for heat carriers are spins and phonons, and which dominates the thermal conductivity is a persistent issue. 
Typically, phonons dominate the longitudinal thermal conductivity while the thermal Hall effect (THE) is primarily associated with spins, which requires time-reversal symmetry breaking.
The coupling between phonons and spins usually depends on spin-orbit interaction and is relatively weak.
Here, we propose a new mechanism for this coupling and the associated THE: the skew scattering of phonons via spin fluctuations by the scalar spin chirality. 
This coupling does not require spin-orbit interaction and is ubiquitous in Mott insulators, leading to a thermal Hall angle on the order of $10^{-3}$ to $10^{-2}$. 
Based on this mechanism, we investigate the THE in YMnO$_3$ with a trimerized triangular lattice where the THE beyond spins was recognized, and predict the THE in the Kagome and square lattices.
\end{abstract}


\maketitle


\section{Introduction}

In Mott insulators, spins play crucial roles in low-energy phenomena, including magnetic, optical, and thermodynamic properties.
Regarding transport properties, the charge current is blocked by the Mott gap while the spin current can be carried by spin excitations. 
The spin system also carries heat current, which can be detected by thermal conductivity and thermal Hall conductivity measurements. 
Additionally, phonons are also active in Mott insulators, contributing to the heat current simultaneously.
Hence, thermal conductivities are discussed in terms of both spins and phonons. 
However, it is often unclear which is dominating the heat transport, necessitating quantitative discussion to resolve this issue. 

Accordingly, the thermal Hall effect (THE) of magnons has been discussed theoretically~\cite{katsura2010theory,matsumoto2011rotational,owerre2016first,owerre2017topological,zhang2019thermal,zhang2024thermal}, and observed experimentally in several materials like a pyrochlore magnet Lu$_2$V$_2$O$_7$~\cite{onose2010observation}, a pyrochlore spin liquid Tb$_2$Ti$_2$O$_7$~\cite{hirschberger2015large}, a Kagome magnet Cu(1,3-benzenedicarboxylate)~\cite{hirschberger2015thermal}, a van der Waals magnet VI$_3$~\cite{zhang2021anomalous}, and a polar magnet GaV$_4$Se$_8$~\cite{akazawa2022topological}.
The half-integer quantized THE was observed from the Majorana fermion in a Kitaev spin liquid~\cite{kasahara2018majorana}. 
On the other hand, theories for the THE of phonons have also been developed~\cite{sheng2006theory,kagan2008anomalous,wang2009phonon,zhang2010topological,agarwalla2011phonon,qin2012berry,saito2019berry} following its discovery on paramagnetic dielectric Tb$_3$Ga$_5$O$_{12}$~\cite{strohm2005phenomenological,inyushkin2007phonon} and Tb$_3$Gd$_5$O$_{12}$~\cite{mori2014origin}, Kitaev spin liquid candidates $\A$-RuCl$_3$~\cite{hentrich2018unusual} and Na$_2$Co$_2$TeO$_6$~\cite{chen2024planar}, a spin ice Pr$_2$Ir$_2$O$_7$~\cite{uehara2022phonon}, the doped high-$T_c$ cuprates~\cite{grissonnanche2019giant,boulanger2020thermal}, another paramagnetic dielectric SrTiO$_3$~\cite{li2020phonon,sim2021sizable}, a topological insulator Bi$_{2-x}$Sb$_x$Te$_{3-y}$Se$_y$~\cite{sharma2024phonon}, an antiferromagnetic insulator Cu$_3$TeO$_6$, and a polar magnet (Zn$_x$Fe$_{1-x}$)Mo$_3$O$_8$~\cite{ideue2017giant}.
Mostly, phonon THE are based on the Raman-type (or spin-phonon) interaction $\vec{M}\cdot \vec{L}$ between the magnetic moment $\vec{M}$ and the phonon angular momentum $\vec{L} = \vec{u}\times\vec{p}$ via the relativistic spin-orbit interaction. 
Here, $\vec{u}$ is the nucleus displacement and $\vec{p}$ is the nucleus momentum.
However, even without magnetic moments $\vec{M}$, the phonon THE can emerge under a magnetic field in both ionic~\cite{agarwalla2011phonon} and neutral atomic crystals~\cite{saito2019berry}.

Given the numerous proposed mechanisms for the THE in Mott insulators, it is crucial to develop a quantitative theory to analyze experimental results and reveal its microscopic mechanisms. 
The multiferroic YMnO$_3$ is an ideal system for this purpose because its spin Hamiltonian was established by various experimental methods, and accurate measurements of its longitudinal and transverse thermal conductivities was achieved~\cite{kim2024thermal}. 
By combining Monte Carlo simulations and numerical solutions of the Landau-Lifshitz-Ginzburg equation, the thermal Hall conductivity ($\kappa_{yx}$) of its spin Hamiltonian has been estimated. 
Theoretical $\kappa_{yx}$ increases toward the Néel temperature ($T_N$) aligning with the development of scalar spin chirality (SSC) fluctuation, but rapidly drops above $T_N$ since the propagation of spin fluctuation is suppressed. Contrary to the semiquantitative agreement between theory and experiment below $T_N$, experimental $\kappa_{yx}$ persists above $T_N$. This suggests that phonons, which can propagate above $T_N$, are influenced by the spin fluctuation with the SSC, contributing to $\kappa_{yx}$. 

\begin{figure*}
    \centering
    \includegraphics[width=1.5\columnwidth]{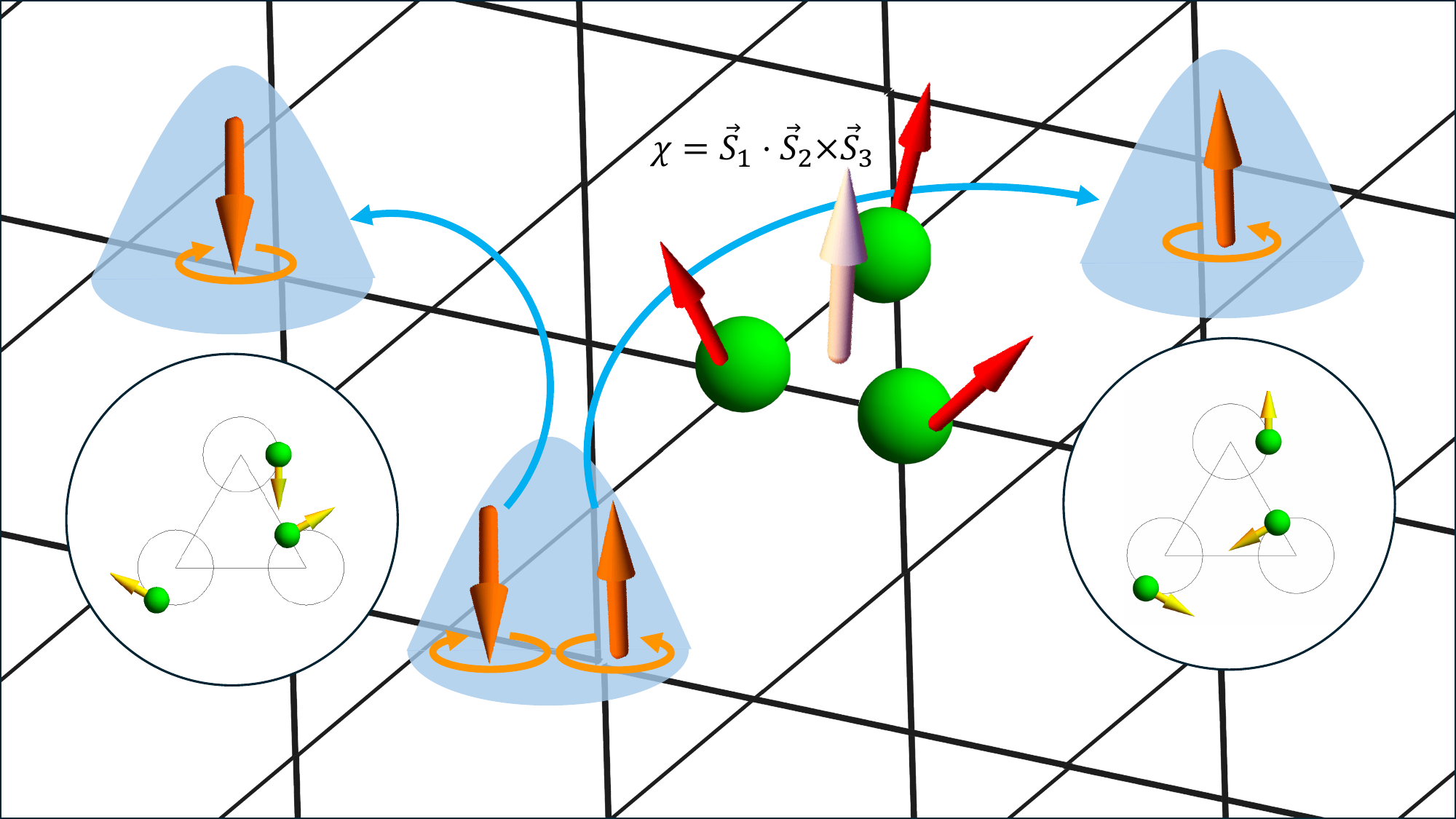}
    \caption{{\bf The phonon skew-scattering by the scalar spin chirality.} When there is the noncoplanar spin structure (or spin fluctuation) of magnetic moments, it develops the scalar spin chirality. The scalar spin chirality modifies the electronic many-body wavefunction and gives rise to the emergent Raman interaction. The Raman interaction scatter the left and right circularly polarized chiral phonons (in the circles) differently, causing the thermal Hall Effect. Here, the red arrows are the spins, the white arrow is the scalar spin chirality, the yellow arrows are the nuclei momentum, the orange arrows are the direction of phonon angular momentum, and the blue arrows are the propagation direction of the phonons.}
    \label{fig:1}
\end{figure*}

Based on this observation, we develop a theory of phonon skew scattering by the spin fluctuation with the SSC in this paper.
We depict the physics in Fig.~\ref{fig:1}. 
We assume a single triangle with a noncoplanar spin structure having a Mott insulating phase without the relativistic spin-orbit coupling. 
Then, the SSC $\chi = \vec{S}_i \cdot \vec{S}_j \times \vec{S}_k$, which characterizes the noncoplanar spin structure~\cite{wen1989chiral,kawamura1992chiral,shindou2001orbital,taguchi2001spin,lee2006doping,kawamura2010chirality,nagaosa2012gauge,nagaosa2012emergent}, is developed, where $\vec{S}_i$ is the spin at $i$th lattice site. 
Although the SSC gives rise to topological Hall Effect for electrons~\cite{bruno2004topological,neubauer2009topological,nagaosa2010anomalous,kanazawa2011large,ishizuka2018spin,ishizuka2021large}, it is very unlikely that the SSC plays a role for phonons which are charge neutral. 
Nevertheless, the SSC modifies the electronic wavefunction in the Mott insulating phase, resulting in the emergent Raman interaction. The emergent Raman interaction causes the antisymmetric scattering between the left and the right circularly polarized chiral phonons. Accordingly, the THE emerges.

The rest of the paper is organized as follows. 
In Sec.~\ref{sec:2}, we briefly explain the Born-Oppenheimer approximation. 
In Sec.~\ref{sec:3}, based on the noncoplanar spin structure in a triangle, we explain how the Raman interaction arises from the electronic wavefunction, and obtain the possible form of the Raman interaction by group theoretical methods, and estimate the magnitude of the effective field of the Raman interaction. 
In Sec.~\ref{sec:4}, we discuss the same as Sec.~\ref{sec:3} in the noncoplanar spin structure of a square.
In Sec.~\ref{sec:5}, we compute the antisymmetric part of the scattering rate, and employ the Boltzmann theory to obtain the thermal Hall conductivity. 
In Sec.~\ref{sec:6}, we provide discussions and draw conclusions.

\section{The Born-Oppenheimer Approximation \label{sec:2}}

The complete Hamiltonian for the solids are given by 
\alg{
H = H_{nu} + H_{el} + V
}
where $H_{nu} = -\sum_a \nabla_{R_a}^2/2M_a$ is the kinetic Hamiltonian for nuclei, $H_{el} = -\sum_i \nabla_{r_i}^2/2m_e$ is that for electrons, and $V = -\sum_{i,a}Z_a/r_{ia} + \sum_{i>j}1/r_{ij} + \sum_{a>b}Z_aZ_b/r_{ab}$ is the summation of Coulomb interactions. 
Here, $\{\V r_i\}$ is the position vectors for electrons, $\{\V R_a\}$ is the position vectors for nuclei, $M_a$ is the mass of $a$th nucleus, $m_e$ is the electron mass, $Z_a$ is the charge of nucleus, and the unit charge $e$ and $\hbar$ are unit.
The conventional Born-Oppenheimer approximation~\cite{oppenheimer1927quantentheorie} estimates the eigenfunction of the Hamiltonian as
\alg{
\ket{\Psi} \approx \ket{\psi_{el}(\{\V r_i\},\{\V R_a\})} \ket{\psi_{nu}(\{\V R_a\})},
}
where
\[
(H_{el}+V)\ket{\psi_{el}} = E_{el}(\{\vec{R}_a\})\ket{\psi_{el}},
\]
and
\[
(H_{nu}+E_{el}(\{\vec{R}_a\}))\ket{\psi_{nu}} = E_{nu}\ket{\psi_{nu}}.
\]

However, this is not a complete theory. When the magnetic field is applied, the nuclei Hamiltonian $H_{BO}^0 = H_{nu}+E_{el}$ changes to
\[
\sum_{a} \f{(\vec{P}_a-Z_a\vec{A}_a)^2}{2M_a} + E_{el}(\{\vec{R}_a\}).
\]
Here, $\vec{A}_a$ is the vector potential applied to $a$-th nucleus due to the external magnetic field. 
This does not reflect the charge neutrality of phonons.
To consider the charge neutrality, we should include the correction term.
Since the length scales $l_{el}$ and $l_{nu}$ for $\ket{\psi_{el}}$ and $\ket{\psi_{nu}}$ is estimated as $l_{el}/l_{nu} = (M_a/m_e)^{1/4}$, the derivative $\grad_{R_a}$ of $\ket{\psi_{el}}$ cannot be neglected.
Accordingly, the Born-Oppenheimer approximation becomes the following~\cite{mead1979determination,mead1992geometric,zhang2010topological,qin2012berry,saito2019berry}:
\alg{
H_{BO} = \sum_{a} \f{(\vec{P}_a-Z_a\vec{A}_a-\vec{a}_a)^2}{2M_a} + E_{el}(\{\vec{R}_a\}).
}
Here, $\vec{a}_a = -i\bra{\psi_{el}}\nabla_{R_a}\ket{\psi_{el}}$ is the Berry connection coming from the electronic many-body wavefunction $\ket{\psi_{el}}$. 
In a single hydrogen-like atom, the Berry connection exactly cancels out the external vector potential, explaining the screening of magnetic field by electrons~\cite{mead1992geometric}. In the lattice, although the field is screened, the cancellation of Berry connection is not perfect, resulting in the finite Raman interaction, phonon Berry curvature, and associated intrinsic THE~\cite{saito2019berry}. This is similar to the Aharonov-Bohm phase.
In the following sections, we show that the SSC also causes the emergent Raman interaction from the electronic many-body wavefunction. 

\section{The emergent Raman interaction in the triangle \label{sec:3}}

\subsection{The model}

\begin{figure}
    \centering
    \includegraphics[width=\columnwidth]{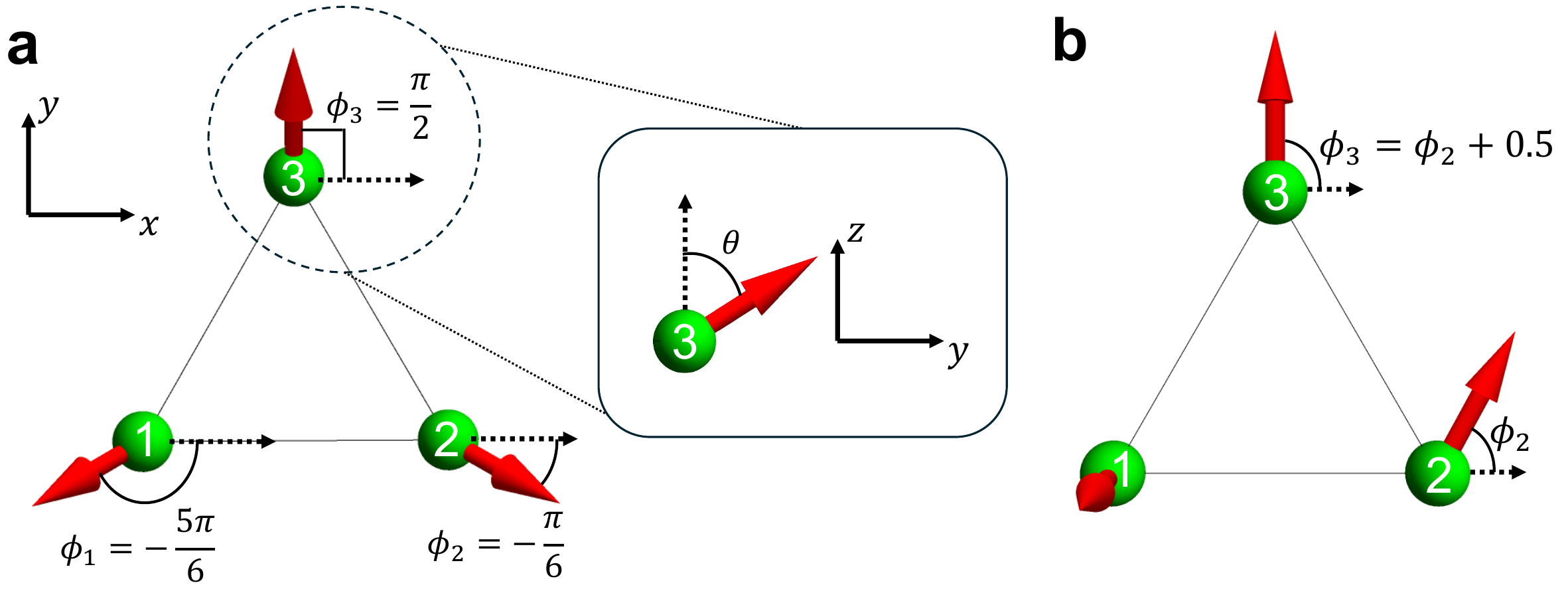}
    \caption{{\bf The spin structure in a triangle.} (a) First, for facilitating the argument, we assume that the noncoplanar spin structure respects the threefold rotation symmetry around the $z$-axis. We set $\vec{S}_i = S(\sin\ta\cos\phi_i,\sin\ta\sin\phi_i,\cos\ta)$, such that $\phi_1 = -5\pi/6, \phi_2 = -\pi/6$, and $\phi_3 =\pi/2$. 
    (b) Next, for investigating the correlation between SSC and emergent field, we assume the symmetry-broken noncoplanar spin structure, $\ta_1 = 0, \ta_2 = \ta_3 = \pi/2, \phi_3=\phi_2 +0.5$ with $\phi_2 \in [0,\pi]$. Here, the SSC does not change with varying $\phi_2$, $\chi\approx0.4794$.}
    \label{fig:2}
\end{figure}

First, we describe the physical situation and the model.
We employ two cases: a single triangle which has three sites ($a=1,2,3$), and a square which has four sites ($a=1,2,3,4$). 
Here, we primarily describe the triangular lattice.
The noncoplanar spin structure in the triangle is assumed to be invariant under threefold spatial and spin rotation about the $z$-axis. Here, the polar angle of each spin is $\ta_1 = \ta_2 = \ta_3 = \ta$, and the azimuthal angles of each site are $\phi_1 = -5\pi/6, \phi_2=-\pi/6$, and $\phi_3 = \pi/2$. 
The spin vector is $\vec{S}_a = S(\sin\ta_a\cos\phi_a, \sin\ta_i\sin\phi_a,\cos\ta_a)$. [See Fig.~\ref{fig:2}.]
The reason to take such a symmetric spin configuration is that (i) it facilitates the analysis, and (ii) although we have asymmetric spin configuration, the symmetry will be recovered after the average over the thermal fluctuations. 
The case of generic spin configurations will be discussed later. [See Fig.~\ref{fig:2}(b).]

To investigate the THE from phonon scattering, we consider the half-filled Mott insulating phase. 
We consider the double exchange model without spin-orbit coupling:~\cite{ye1999berry,hamamoto2015quantized}
\begin{align}
    H_{el} = H_t + H_J, \label{eq:4}
\end{align}
where
\[
H_t = \sum_{ij,\A} t_{ij}c_{i\A}^\dg c_{j\A},
\]
is the kinetic Hamiltonian, and
\[
H_J = -J_H \sum_{i,\A\B} \vec{S}_i \cdot c_{i\A}^\dg \vec{\ma}_{\A\B} c_{i\B},
\]
is the double exchange ($J_H>0$). 

We transform the coordinate that the local $z$-axis at each site points to $\vec{S}_i$ direction. That is,
\alg{
H_J = -J_H \sum_{i,\A\B} c_{i\A}^\dg \ma^z_{\A\B}c_{i\B}.
}
We employ $S=1$.

Furthermore, we decompose $t_{ij} = t+\dt t_{ij}$.
Here, $t$ is the transfer integral when the nuclei are at their equilibrium position, and $\dt t_{ij}$ is the change of the transfer integral due to the displacement of nuclei. 
Here, we assume that $\dt t_{ij}$ only depends on the distance between $i$ and $j$ sites.
Accordingly, $H_t$ is decomposed into $H_{t0}$ and $\dt H_{t}$. $H_{t0}$ is the kinetic Hamiltonian when all nuclei are at the equilibrium, and $\dt H_t$ is the change of kinetic Hamiltonian by the nuclei displacement. Notably, the model has two control parameters: the polar angle $\ta$ and the equilibrium transfer integral $t$.

\subsection{The Berry connection and the emergent field \label{sec:3-1}}

\begin{figure}
    \centering
    \includegraphics[width=\columnwidth]{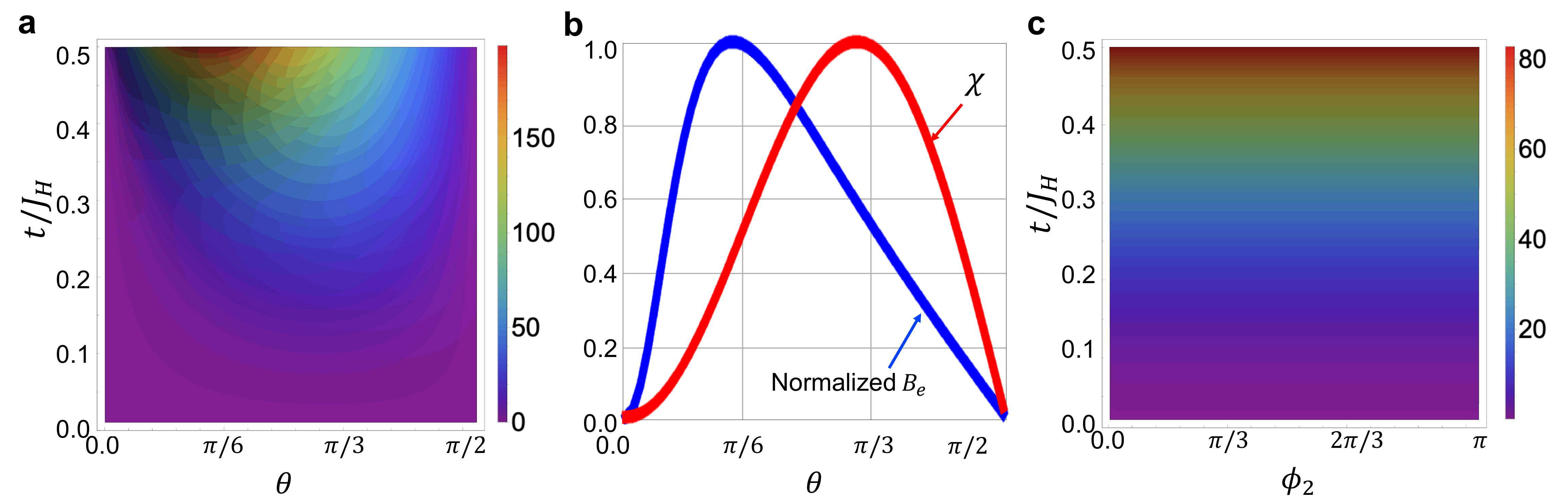}
    \caption{{\bf The strength of emergent field in a triangle from SSC.} (a) The strength of emergent field (unit $T$) as a function of $\ta$ and $t/J_H$ from the spin configuration in Fig.~\ref{fig:2}(a). (b) The SSC $\chi = \vec{S}_1 \cdot \vec{S}_2 \times \vec{S}_3$ and the normalized emergent field at $t=0.4$ as a function of $\ta$. They are qualitatively identical, in that they vanishes at $\ta =0,\pi/2$.
    (c) The field strength $B_e$ as a function of $\phi_2$ and $t/J_H$ from the spin configuration in Fig.~\ref{fig:2}(b). Although the spin structure changes, the field strength is constant. This supports that the emergent field is correlated with the SSC.}
    \label{fig:3}
\end{figure}

We acquire the Berry connection by the SSC by employing the following way.
First, we calculate the second-order perturbed single-body eigenstates $\ket{\varphi_i(t,\theta,\{\dt t_{ij}\})}$ whose energy is $E_i$ from $H_0 =H_{t0}+H_J$ and $V = \dt H_t$ ($i=1,...,6$).
The sequence of energy is $E_1 \leq E_2 \leq ... \leq E_6$.
Then, the ground state many-body wavefunction is 
\[
\ket{\det{\varphi_{1}\varphi_{2}\varphi_{3}}} = \f{1}{\sqrt{3!}}\ep_{ijk}\ket{\varphi_i}\otimes \ket{\varphi_j}\otimes \ket{\varphi_k}.
\]
Here, $\ep_{ijk}$ is the antisymmetric tensor.
Then, one can simply show that the Berry connection becomes
\alg{
\vec{a}_a =& -i\bra{\det{\varphi_{1}\varphi_{2}\varphi_{3}}} \p_{R_a} \ket{\det{\varphi_{1}\varphi_{2}\varphi_{3}}}
\nq \sum_{(ij)} (\p_{R_a}\dt t_{ij})C_{ij}. \nonumber
}
Here, $ C_{ij}\equiv\sum_k (-i) \bra{\varphi_k}\p_{\dt t_{ij}}\ket{\varphi_k}$, and $(ij) = (12),(23),(31)$.
By the expansion of $C_{ij}$ up to the lowest order of $\dt_{ij}$, we obtain
\begin{align}
C_{12} =& A (\dt t_{23} - \dt t_{31}),\n
C_{23} =& A (\dt t_{31} - \dt t_{12}),\n
C_{31} =& A (\dt t_{12} - \dt t_{23}).\nonumber
\end{align}
Here, $A = A(t/J_H,\ta)$ is the coefficient varying with $t/J_H$ and $\ta$. 
The coefficients of $C_{ij}$ are the same due to the threefold rotation symmetry.

We can approximate further by $\dt t_{ij}(\vec{R}_a) \approx \dt t_{12}(\vec{R}_a^0) + \vec{u}_a \cdot \nabla_{R_a} \dt t_{ij}|_0 = \vec{u}_a \cdot \nabla_{R_a} \dt t_{ij}|_0$, where $\vec{R}_a = \vec{R}_a^0 + \vec{u}_a$, $\vec{R}_a^0$ is the equilibrium position of $a$-th nucleus, and $\vec{u}_a$ is the displacement. The Berry connection is now
\alg{
\vec{a}_a \approx& A \sum_{(ijk)} (\nabla_{R_a} \dt t_{ij})[(\nabla_{R_b} \dt t_{jk} - \nabla_{R_{b}} \dt t_{ki})]\cdot \vec{u}_b. \label{eq:6}
}
Here, $\sum_{(ijk)} = \sum_{(ij)}\sum_{k\neq i,j}$.
Since the emergent field from the Berry connection is $\vec{B}_a = \nabla_{R_a} \times \vec{a}_a$, the field strength is estimated as $B_{e} \equiv |\vec{B}_a| \sim  A(t/J_H,\ta)(\nabla_{R_a} \dt t_{ij})^2$.
Notably, the typical value of $J_H \sim 1~eV$, $A \sim 1~ eV^{-2}$, and $|\nabla_{R_a}\dt t_{ij}| \sim 100~meV/\AA$~\cite{coropceanu2007charge}, so the unit of field strength is $\sim 6.50 \times 10^2 ~T$.

In Figs.~\ref{fig:3}(c) and (d), we showcase the estimated field strength $B_e$ in the plane of $t/J_H$ and $\ta$ in the unit of tesla, and compare with the SSC $\chi$. 
The field strength vanishes at $\ta = 0,\pi/2$, is maximized about $\ta_{max} \approx 0.5$, and is increasing with increasing $t$. 
The $t$ dependence changes as $\ta$ changes.
The field strength is reaches at most $\sim 200~T$ ($\sim 10~meV$) in the parameter range, which is enough to induce the observable THE.
Although the SSC is maximum at $\ta = \arccos(-1/3)/2 \approx 0.955$, the SSC and the field strength are qualitatively identical, in that both vanish at $\ta=0,\pi/2$. 

In addition, to confirm that the SSC and emergent field are correlated, we acquire the field strength when the spin configuration changes while the SSC is invariant. 
We obtain the emergent field for the symmetry-broken configuration in Fig.~\ref{fig:2}(b). 
$\vec{S}_1 \parallel \hat z$, $\vec{S}_2$ and $\vec{S}_3$ is in the $xy$-plane. 
We rotate $\vec{S}_2$ and $\vec{S}_3$ while fixing the angle between two vectors as $0.5~rad$. 
Namely, we set $\ta_1 = 0, \ta_2 = \ta_3 = \pi/2, \phi_3 = \phi_2 + 0.5$ with $\phi_2 \in [0,\pi]$.
As the rotation of spins is in $SO(3)$, 
the SSC is constant as $\chi\approx 0.4794$ despite the rotation. 

Two important facts should be noted. 
First, the form of the emergent Berry connection is the same as Eq.~\ref{eq:6}, which means that the threefold rotation symmetry is restored by the thermal fluctuation. 
This implies the $SO(3)$ invariance of emergent field, identical to the SSC.
Next, the associated emergent field in the plane of $t/J_H$ and $\phi_2$ is shown in Fig.~\ref{fig:3}(c). 
The emergent field strength only varies by $t/J_H$ while remains constant by the rotation angle $\phi_2$. These facts imply that the emergent field is correlated with the SSC.

\subsection{The possible form of Raman interaction \label{sec:3-2}}

\begin{figure*}[t]
    \centering
    \includegraphics[width=2\columnwidth]{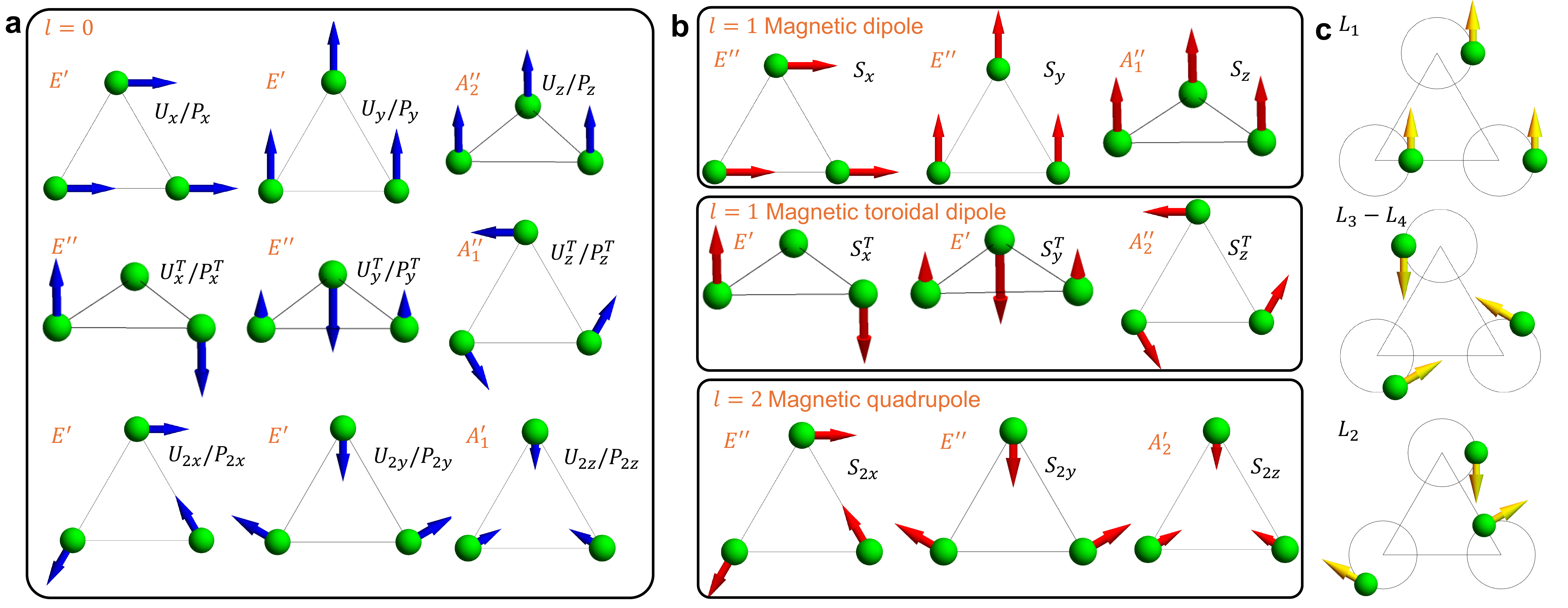}
    \caption{{\bf The nuclei displacement/momentum, spin configurations, and chiral phonon modes in the triangle.} The classification of (a) the nuclei displacement and momentum, and (b) the spin configuration into the IRREPs of the spin group $G = (SO(3)\times D_{3h}) +\tau(SO(3)\times D_{3h})$. Notably, all modes of the nuclei displacement and momentum are in $l=0$ IRREPs while the spin configurations are in $l\neq 0$ IRREPs. Therefore, the spin and nuclei can couple only by the SSC $\chi$, which is in the $(l=0) \times A_1''$ IRREP. (c) The chiral phonon modes, which are the eigenstates of $C_{3z}$ symmetry. These phonon modes are the candidates that can couple to the SSC. However, for $L_1$ and $L_3-L_4$, since $\dt t_{12}=\dt t_{23} = \dt t_{31}$, $\vec{a}_a = 0$. $L_2$ is the only chiral phonon that couples to the SSC. 
    The blue arrows are either nuclei displacement or momentum, the red arrows are the spins, and the yellow arrows are the nuclei momentum. Note that the nuclei displacement and momentum modes are classified into the same way. The length difference of the arrows within each mode indicates the magnitude difference.}
    \label{fig:4}
\end{figure*}

In Eq.~\ref{eq:6}, we acquire the Berry connection from the electronic many-body wavefunction. 
However, the Berry connection can vanish by symmetries. 
For example, the atom 1 in Fig.~\ref{fig:2} moves away from its equilibrium position respecting $\dt t_{12} = \dt t_{31}$, $\vec{a}_1 = 0$. 
Accordingly, when $\dt t_{12}=\dt t_{23} = \dt t_{31}$, $\vec{a}_a = 0$ for all $a$.
Therefore, to find the condition when the Berry connection can be finite, we here induce the general form of the Raman interaction by the group theoretical methods.


Here, the system does not have spin-orbit coupling, so we should consider the spin group~\cite{litvin1977spin,vsmejkal2022emerging,schiff2023spin}.
The total symmetry group of a single triangle is $G = \mathcal G + \tau \mathcal G$, where  $\mathcal{G} = SO(3)\times G_p$, $\tau$ is the time-reversal symmetry. $SO(3)$ stands for the global rotation applied only for the spin sector, and $G_p$ is the point group of the system for the spatial sector.
The point group of a single triangle is $G_p = D_{3h}$, containing elements such as $I$, $C_{3z}$, $C_{3z}P$, $C_{2}'$, $C_{2}'P$, $M$.
Here, $I$ is the identity, $C_{n}$ is the $n$-fold rotation, $P$ is the inversion, and $M$ is the mirror.
The IRREPs of $G$ is described by $d^l(SO(3)) \times d^\nu(D_{3h})$, where $l$ denotes the angular momentum of spin part and $\nu$ is for the IRREPs of $D_{3h}$~\cite{aroyo2006bilbao,aroyo2006bilbao2,aroyo2011crystallography}.

In Figs.~\ref{fig:4}(a) and (b), we showcase all possible nuclei displacement and momentum modes, and spin configuration classified into the IRREPs.
Especially, the spin configuration can be classified by a canonical method called  cluster multipole expansion, by which the angular momentum of the spin configuration can be obtained~\cite{suzuki2017cluster,oh2018magnetic,suzuki2019multipole,oh2023transverse}. 
The nuclei displacement and momentum is in the $l=0$ IRREPs since they are all invariant under global spin rotation. 
However, the spin configuration is in $l\neq 0$ IRREPs since they are not invariant under global spin rotation. 
This implies that the spin configuration cannot directly couple to the nuclei displacement and momentum without spin-orbit coupling, as anticipated from the global spin rotation symmetry.

Nevertheless, interestingly, the SSC $\chi = \vec{S}_1 \cdot \vec{S}_2\times \vec{S}_3$ is in the $(l=0) \times A_1''$ IRREP since this is a scalar which is invariant under $SO(3)$, so this can couple to the nuclei displacement and momentum. The possible Raman interaction is written as $(H_R)_i = \B_i \chi L_i$, where $\B_i$ are constants and
\alg{
L_1 =& U_xP_y - U_yP_x, \n
L_2 =& U_{2x}P_{2y}- U_{2y}P_{2x}, \n
L_3 =& U_z^T P_{2z}, L_4 = U_{2z}P_{z}^T, \n
L_5 =& U_xP_{2y}-U_yP_{2x}, \n
L_6 =& U_{2x}P_y - U_{2y}P_x, \n
L_7 =& U_{x}^T P_y^T - U_y^T P_{x}^T.
}
$U_x, U_y, U_{2x}, U_{2y}, U_{2z}, U_x^T, U_y^T, U_z^T$ are the nuclei displacement modes, and $P_x, P_y, P_{2x}, P_{2y},  P_{2z}, P_x^T, P_y^T, P_z^T$ are the nuclei momentum modes defined in Fig.~\ref{fig:4}(a). 
It is note-worthy that the modes of nuclei displacement and momentum are identical since both of them are proper vectors.

One can analyze the nuclei motion related to each $L_i$ by the following procedure. For instance, one can let $U_x = U \cos\zeta$, $U_y = U\sin\zeta$, $P_x = -P\sin\zeta, P_y=P\cos\zeta$, and zero otherwise. 
Then, $L_1 = UP$ while other $L_i$ are zero. 
When drawing the nuclei position and momentum, we confirm that $L_1$ corresponds to the rotation of each atom around its equilibrium position, in which the rotations are in phase as we depicted in Fig.~\ref{fig:4}(c).
By this procedure, we find that $L_1, L_2, L_3-L_4$ are the rotation of the nuclei surrounding its equilibrium positions, which are the eigenmodes of $C_{3z}$ symmetry.
These are related to the chiral phonons~\cite{zhang2015chiral,chen2019chiral,park2020phonon}.
$L_5, L_6$ are unphysical motions, and $L_7$ is the vibration along $z$-direction. 

The rotation induces the angular momentum that can be naturally coupled to the emergent field from the Berry connection. 
Thus, $L_1, L_2, L_3-L_4$ are the candidates for the Raman interaction. 
However, the Berry connection vanishes in $L_1$ and $L_3-L_4$ since these keep the symmetry $\dt t_{12} = \dt t_{23} = \dt t_{31}$. 
Hence, the only possible form of Raman interaction is $H_R = \B \chi L_2/M = -(P_{2x}a_{2x} + P_{2y}a_{2y})/M$ with
\alg{
a_{2x} = \B\chi U_{2y},~~ a_{2y} = -\B\chi U_{2x}.
}
If we transform $a_{2x}$ and $a_{2y}$ into the atomic basis,
\alg{
\vec{a}_1 =& \f{1}{2}\vec{B}_{\text{eff}} \times (2\vec{u}_1 - \vec{u}_2 - \vec{u}_3) , \n
\vec{a}_2 =& \f{1}{2}\vec{B}_{\text{eff}} \times (2\vec{u}_2 - \vec{u}_3 - \vec{u}_1) , \n
\vec{a}_3 =& \f{1}{2}\vec{B}_{\text{eff}} \times (2\vec{u}_3 - \vec{u}_1 - \vec{u}_2),
}
up to the gauge transform.
Here, $\vec{B}_{\text{eff}} = \B\chi\hat z/3 \sim B_e \hat z $. Thus, now we have the Hamiltonian for the nuclei motion in 2D of a single triangle with harmonic approximation.
\alg{
H_{nu} = \sum_a \f{(\vec{P}_a - \vec{a}_a)^2}{2M} + \sum_{ab,mn} u_a^m D_{ab}^{mn} u_b^{n},
}
with $a,b = 1,2,3$ and $m,n=x,y$.

To consider the phonon, we expand our attention from a single triangle to the lattice system. Accordingly, the emergent Raman interaction at the $i$-th unit cell is:
\alg{
H_R(r) =& -\sum_{a} \f{\vec{P}_a \cdot \vec{a}_a}{M} \dt(r-\vec{R}_{i,a}) 
\nq -\f{B_e}{2M}\sum_{ab}   \Lambda_{ab}L_{ba}^z\dt(\vec{r}-\vec{R}_{i,a}) .
\label{eq:15}
}
where $\Lambda_{ab} = 3\dt_{ab}-1$, $\vec{L}_{ba} = \vec{u}_b\times \vec{P}_a$ is the generalized phonon angular momentum, $\vec{R}_{i,a} = \vec{R}_i+\vec{d}_a$, $\vec{R}_i$ is the unit cell position, and $\vec{d}_a$ is the sublattice position in the unit cell. 

Closing this section, we note several points. 
First, the conventional phonon angular momentum is defined as $\vec{L} = \sum_a \vec{L}_{aa}$~\cite{park2020phonon}, so we call $\vec{L}_{ba}$ the generalized phonon angular momentum.
Second, the total field applied to $a$-th sublattice is $\sum_b\Lambda_{ab}B_e\hat z = 0$, which means the absence of external magnetic field~\cite{saito2019berry}. 
However, although the field is canceled, the Berry connection remains, similar to the Aharonov-Bohm Effect. This is responsible for the phonon skew-scattering in Sec.~\ref{sec:5}.

\section{The emergent Raman interaction in the square lattice \label{sec:4}}

\subsection{The model}

\begin{figure}[t]
    \centering
    \includegraphics[width=\columnwidth]{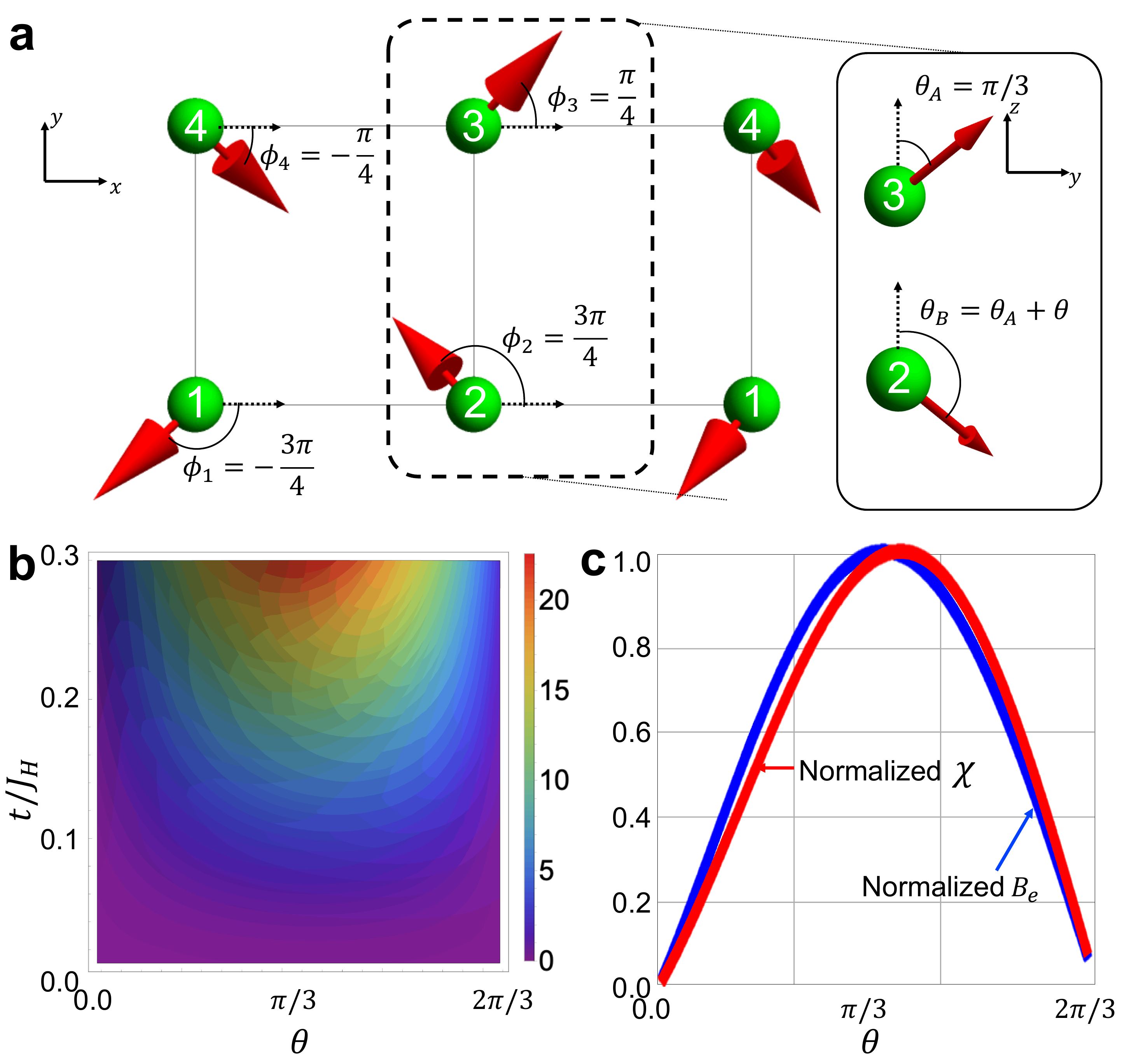}
    \caption{{\bf Emergent field from the SSC in a square lattice.} (a) We assume the noncoplanar spin structure in two neighboring squares. 
    We set $\vec{S}_i = S(\sin\ta_i\cos\phi_i,\sin\ta_i\sin\phi_i,\cos\ta_i)$, such that $\ta_1=\ta_3 = \ta_A$, $\ta_2=\ta_4=\ta_B=\ta_A+\ta$, $\phi_1 = -3\pi/4, \phi_2 = 3\pi/4$, $\phi_3 = \pi/4$, and $\phi_4 =-\pi/4$. (b) The strength of emergent field (unit $T$) as a function of $\ta$ and $t/J_H$. (c) The normalized SSC $\chi = \vec{S}_1 \cdot \vec{S}_2 \times \vec{S}_3$ and the normalized emergent field at $t=0.3$ as a function of $\ta$. }
    \label{fig:5}
\end{figure}

Here, we briefly discuss the square lattice case based on the discussion in the previous section.
For the same purpose as previous section, the noncoplanar spin structure in the square is assumed as Fig.~\ref{fig:5}(a).
In two neighboring squares, there are total six spins, where the left two and the right two spins are identical. 
The spin is $\vec{S}_i = (\sin\ta_i\cos\phi_i, \sin\ta_i\sin\phi_i,\cos\ta_i)$.
The polar angle of each spin is $\ta_1 = \ta_3 = \ta_A, \ta_2 = \ta_4 = \ta_B, \phi_1 = -3\pi/4, \phi_2 = 3\pi/4, \phi_3 = \pi/4$, and $\phi_4 = -\pi/4$.
The Hamiltonian is the same as Eq.~\ref{eq:4}.
We fix $\ta_A$ to be $\pi/3$, and change $t \in [0,0.3]$ and $\ta_B = \ta_A + \ta$, $\ta \in [0,2\pi/3]$.
[See Fig.~\ref{fig:5}(a).]

We can acquire the total SSC of a square by dividing each square into two triangles and adding each SSC up. 
The SSC is given by $\chi = 4(\cos\ta_A - \cos\ta_B) \sin\ta_A\sin\ta_B$. 
This vanishes when either $\ta_A = \ta_B$ , $\ta_A = 0,\pi$, or $\ta_B = 0,\pi$.

\subsection{The Berry connection and the emergent field}

We obtain the Berry connection by the SSC by employing the same method as the previous section.
Here, it is sufficient to consider the left square only since the thermal fluctuation takes the average of the two squares.
The second-order perturbed single-body eigenstates  $\ket{\varphi_i(t,\theta_A,\theta_B,\{\dt t_{ij}\})}$ whose energy is $E_i$ are computed from $H_0 =H_{t0}+H_J$ and $V = \dt H_t$ ($i=1,...,8$).
Again, $H_{t0}$ is the kinetic Hamiltonian when all nuclei are equilibrium, $H_J$ is the double exchange,  $\dt H_t$ is the change of kinetic Hamiltonian by the displacement of nuclei.
Also, the sequence of energy is $E_1 \leq E_2 \leq ... \leq E_8$.
Then, the ground state many-body wavefunction is \[\ket{\det{\varphi_{1}\varphi_{2}\varphi_{3}\varphi_{4}}} = \f{1}{\sqrt{4!}}\ep_{ijkl}\ket{\varphi_i}\otimes \ket{\varphi_j}\otimes \ket{\varphi_k}\otimes \ket{\varphi_l}.
\]
Here, $\ep_{ijkl}$ is the antisymmetric tensor.
Then, one can simply show that the Berry connection becomes
\alg{
\vec{a}_a =& -i\bra{\det{\varphi_{1}\varphi_{2}\varphi_{3}\varphi_{4}}} \p_{R_a} \ket{\det{\varphi_{1}\varphi_{2}\varphi_{3}\varphi_{4}}}
\nq \sum_{(ij)} (\p_{R_a}\dt t_{ij})\mathcal{C}_{ij}. \nonumber
}
Here, $ \mathcal{C}_{ij}\equiv\sum_k (-i) \bra{\varphi_k}\p_{\dt t_{ij}}\ket{\varphi_k}$, and $(ij) = (12),(23),(34),(41)$.
By the expansion of $\mathcal{C}_{ij}$ up to the lowest order of $\dt_{ij}$, we obtain
\[
\mathcal{C}_{ij} = (-1)^b [A_1 \dt t_{jk} + A_2 \dt t_{li}],
\]
where $k,l \neq i,j$. $A_1(t/J_H,\ta), A_2(t/J_H,\ta)$ are the coefficients depending on $t/J_H$ and $\ta$. Thus,
\alg{
\vec{a}_a = \sum_{(ijkl)} (\p_{R_a}\dt t_{ij})(-1)^j [A_1 \dt t_{jk} + A_2 \dt t_{li}].
}
Here, $(ijkl) = (1234),(2314),(3412),(4132)$. When one approximates $t_{ij} \approx \p_{R_b} t_{ij} \cdot \vec{u}_b$, the Berry connection is now
\alg{
\vec{a}_a =& \sum_{(ijkl)} (\p_{R_a}\dt t_{ij})(-1)^j \nm\times[A_1 \p_{R_b}\dt  t_{jk} + A_2 \p_{R_b}\dt  t_{li}] \cdot \vec{u}_b.
}
Hence, the emergent field is approximately $B_e\sim (A_1(t/J_H,\ta)+A_2(t/J_H,\ta))(\p_{R_a}\dt t_{ij})^2$.
We present the emergent field strength in the plane of $t/J_H$ and $\ta$ in Fig.~\ref{fig:5}(b), and compare it with the SSC in Fig.~\ref{fig:5}(c). Similar to Fig.~\ref{fig:3}(d), they are qualitatively identical to each other, in that they vanish $\ta_A = \ta_B =\pi/3$ ($\ta=0$) and $\ta_B = \pi$.

\subsection{The possible form of Raman interaction}

\begin{figure*}[t]
    \centering
    \includegraphics[width=2\columnwidth]{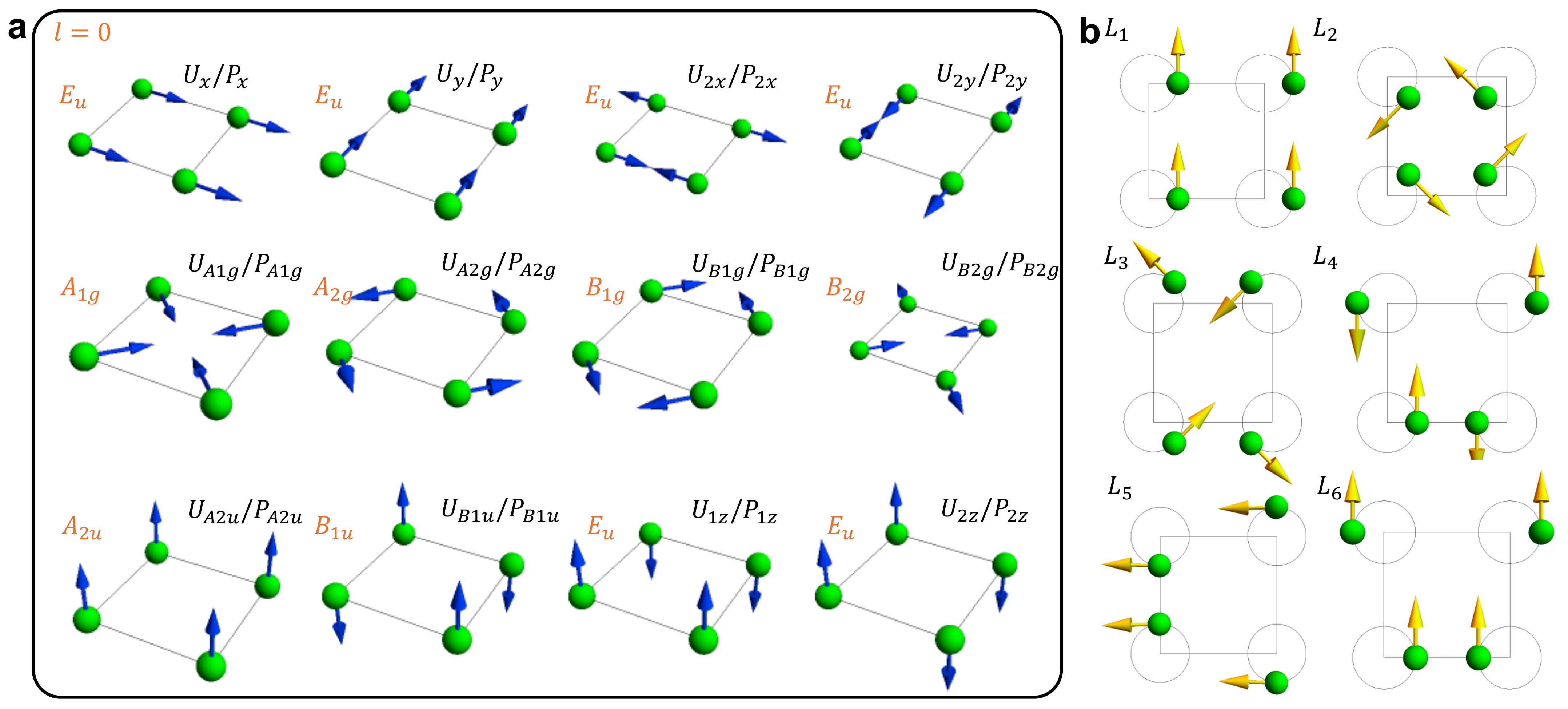}
    \caption{{\bf The nuclei displacement/momentum and chiral phonon modes in the square.} The classification of (a) the nuclei displacement and momentum into the IRREPs of the spin group $G = (SO(3)\times D_{4h}) +\tau(SO(3)\times D_{4h})$. The SSC $\chi$ is in the $(l=0) \times A_{2g}$ IRREP. (b) The chiral phonon modes. $L_1,L_2,L_3,L_4$ are eigenmodes of $C_{4z}$ while $L_5,L_6$ are eigenmodes of $C_{2z}$. 
    $L_3, L_4, L_5, L_6$ can be coupled to the SSC.
    The blue arrows are either nuclei displacement or momentum and the yellow arrows are the nuclei momentum. Note that the nuclei displacement and momentum modes are classified into the same way.}
    \label{fig:6}
\end{figure*}


Similar to the previous case, $\vec{a}_a$ vanishes when $\dt t_{12}=\dt t_{23}=\dt t_{34}=\dt t_{41}$. 
Therefore, we should find the possible form of Raman interaction by symmetry simiarly to the previous section.
Here, the spin point group of square lattice is now $G = (SO(3)\times D_{4h}) +\tau(SO(3)\times D_{4h})$. The IRREPs of the group is $d^l(SO(3)) \times d^\nu(D_{4h})$, where $l$ denotes the angular momentum of spin part and $\nu$ denotes the IRREPs of $D_{4h}$~\cite{aroyo2006bilbao,aroyo2006bilbao2,aroyo2011crystallography}.

By using the cluster multipole expansion method once again, we classify all possible nuclei displacement and momentum modes in Fig.~\ref{fig:6}(a). The SSC is in $(l=0)\times A_{2g}$ IRREP. Accordingly, the possible Raman interactions are $(H_R)_i = \B_i \chi L_i$, where $\B_i$ are constants and 
\alg{
L_1 =& U_xP_y - U_yP_x, \n
L_2 =& U_{A1g}P_{A2g} - U_{A2g}P_{A1g}, \n
L_3 =& U_{B1g}P_{B2g} - U_{B2g}P_{B1g}, \n
L_4 =& U_{2x}P_{2y}-U_{2y}P_{2x}, \n
L_5 =& U_xP_{2y} - U_{2y}P_{x}, \n
L_6 =& U_{2x}P_y - U_{y}P_{2x}.
}
Here, $U_x,U_{A1g},U_{A2g},U_{B1g},U_{B2g},U_{2x},U_{2y}$ are the nuclei displacement modes, and $P_x,P_{A1g} ,P_{A2g},P_{B1g},$ $P_{B2g},P_{2x},P_{2y}$ are the nuclei momentum modes in Fig.~\ref{fig:6}(a). 
The nuclei displacement and momentum modes are identical here as well.

We present the chiral phonons in Fig.~\ref{fig:6}(b) with the same method as the previous section. Since the nuclei in $L_1$ and $L_2$ rotate by keeping $\dt t_{12}=\dt t_{23} = \dt t_{34} = \dt t_{41}$, $L_3,L_4,L_5$, and $L_6$ can be coupled to the emergent field from the SSC.

\section{Thermal Hall Effect from phonon skew-scattering \label{sec:5}}

\subsection{The phonons in the lattice \label{sec:4-1}}

Here, we consider the simplest examples: the trimerized triangular and Kagome lattices, whose unit cell is a triangle. 
Their phonon energy bands are shown in Figs.~\ref{fig:7}(a) and (e), respectively.
Notably, the trimerized triangular lattice is for YMnO$_3$, which showed the THE that cannot be explained only by spins.
Here, we let the lattice constant $a$. 
For the trimerized triangular lattice, we set the intracell longitudinal spring constant $Q_L$ to be $50~meV/a^2$, the intracell transverse spring constant $Q_T$ to be $25~meV/a^2$, the intercell longitudinal spring constant $Q_L'$ to be $25~meV/a^2$, and the intercell transverse spring constant $Q_T'$ to be $12.5~meV/a^2$.
These values are based on the values used in the {\it ab-initio} phonon spectrum computation in YMnO$_3$~\cite{rushchanskii2012ab}.
For the Kagome lattice, we assume no breathing, and set the longitudinal spring constant $Q_L$ to be a typical value of $50~meV/a^2$, and the transverse spring constant $Q_T$ as $15~meV/a^2$.

\subsection{Boltzmann theory \label{sec:4-2}}

In this section, we are going to obtain the skew-scattering by means of the Boltzmann theory. 
We explain how to compute the skew-scattering contribution of THE by the Boltzmann theory ~\cite{leroux1972contribution,ashcroft1976solid,sinitsyn2007anomalous,sinitsyn2007semiclassical,nagaosa2010anomalous,ishizuka2017noncommutative,ishizuka2018spin}.

Let us consider the wave packet made by the eigenfunctions of band $n$, whose position center is at $\vec{r}$ and canonical momentum center is $\vec{k}$. 
The Boltzmann equation is given by 
\alg{
\f{\p f_l}{\p t} + \dot{\vec{r}} \cdot \grad_r f_l + \dot{\vec{p}}\cdot \grad_p f_l = (\f{df_l}{dt})_{coll}.
}
$f_l$ is the nonequilibrium distribution of quasiparticle, and $l=(n,\vec{k})$ is the index for the band and momentum center of the wave packet.
In general, the collision integral is given by
\[
(\f{df_l}{dt})_{coll} = -\sum_{l'} [\w{l'l} f_l(1-f_{l'}) - w_{ll'} f_{l'}(1-f_l)].
\]
Here $\w{ll'}$ means the scattering rate from $l'$ to $l$. 
In the skew-scattering, we only consider the elastic scattering. This reduces the collision integral to
\[
(\f{df_l}{dt})_{coll} =  -\sum_{l'} \{\w{l'l}f_l-\w{ll'}f_{l'}\}.
\]
Thus, the Boltzmann equation for phonons only with temperature gradient is 
\alg{
\f{\p \ep_l}{\p \vec{k}} \cdot \nabla_r T \f{\p f_l}{\p T} = -\sum_{l'} \{\w{l'l}f_l-\w{ll'}f_{l'}\}.
}
Here, $\ep_l$ is the energy dispersion of the phonon.

The scattering rate is decomposed into  symmetric and antisymmetric parts, i.e., $\w{ll'} = \w{ll'}^s + \w{ll'}^a$, where $\w{ll'}^s = (\w{ll'}+\w{l'l})/2, \w{ll'}^a = (\w{ll'}-\w{l'l})/2$. 
Then,
\[
\f{\p \ep_l}{\p \vec{k}} \cdot \nabla_r T \f{\p f_l}{\p T} = -\sum_{l'} [\w{ll'}^s\{f_l-f_{l'}\} - \w{ll'}^a\{f_l+f_l'\}].
\]
We here adopt the relaxation time approximation for the symmetric parts, i.e. $\sum_{l'} \w{ll'}^s\{f_l-f_l'\} = (f_l-f_l^0)/\tau = g_l/\tau$. Here $f_l^0$ is the equilibrium distribution of the quasiparticle at $l$ and $g_l$ is the deviation. Also, $\sum_{l'} \w{ll'}^a f_l$ is absorbed into the relaxation time approximation. Then, the equation becomes:
\alg{
\f{\p \ep_l}{\p \vec{k}} \cdot \nabla_r T \f{\p f_l^0}{\p T} = -\f{g_l}{\tau}
- \sum_{l'} \w{ll'}^ag_l'. \label{eq:18}
}
Here, $\f{1}{\tau} \gg |\w{ll'}^a|$ since $\f{1}{\tau}$ comes from the leading order and $\w{ll'}^a$ comes from the subleading order of impurity strength. We discuss how to acquire $\w{ll'}^a$ in the next section.

Then, we divide $g_l = h_l + \dt h_l$. Also, the temperature gradient is applied along $x$ direction: $\grad_r T = \Delta T \hat x$. Then, 
\alg{
h_l =& - \tau \f{\p \ep_l}{\p k_x}\Delta T \f{\p f_l^0}{\p T},\n
\dt h_l =& - \tau \sum_{l'}\w{ll'}^a h_{l'} =\tau^2 \sum_{l'} \w{ll'}^a \f{\p \ep_{l'}}{\p k_x} \Delta T \f{\p f_{l'}^0}{\p T}.
}
The thermal conductivities are
\alg{
\kappa_{xx} = -\sum_l \ep_l \f{\p \ep_l }{\p k_x} \f{h_l}{\Delta T},~
\kappa_{yx} = -\sum_l \ep_l \f{\p \ep_l }{\p k_y} \f{\dt h_l}{\Delta T}.
}

The trimerized triangular and Kagome lattices are two-dimensional, so we consider the two acoustic bands only. 
Also, we ignore the Umklapp process since we consider $T\ll \Theta_D$. Here, $\Theta_D$ is the Debye temperature. 
For YMnO$_3$, the THE appears near $T_N \approx 72~K$ while $\Theta_D \approx 425~K$~\cite{kim2024thermal,tachibana2005heat}.
We set $\tau \sim  10^{-12}~s$ in the calculation.


\subsection{Scattering rate \label{sec:4-3}}

\begin{figure*}[ht]
    \centering
    \includegraphics[width=2\columnwidth]{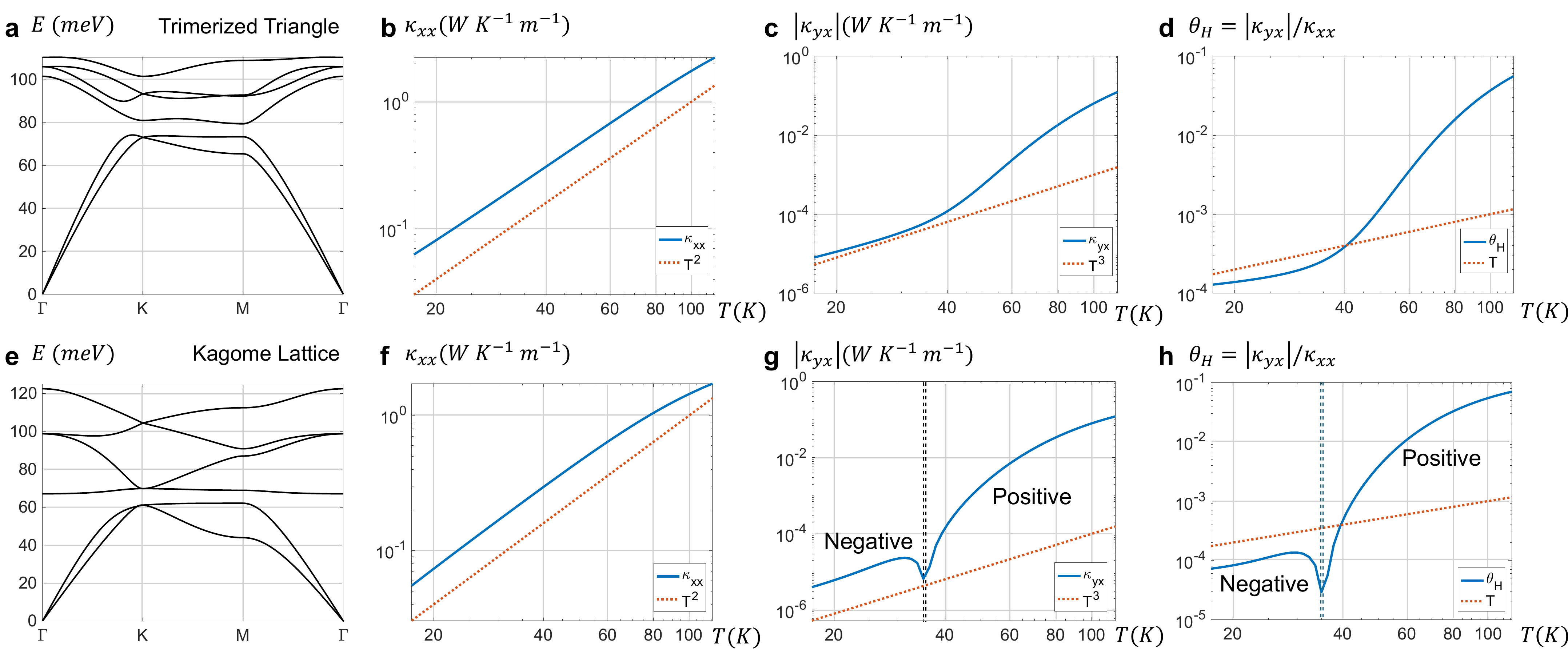}
    \caption{{\bf The THE from the phonon skew-scattering by the SSC.} 
    (a) The phonon spectrum of trimerized triangular lattice with the intracell longitudinal spring constant $Q_L = 50~meV/a^2$, the intracell transverse spring constant $Q_T = 0.5 Q_L$, the intercell longitudinal spring constant $Q_L'=0.5 Q_L$, and the intercell transverse spring constant $Q_T'=0.5Q_L'$. Here $a$ is the lattice constant. The numerical computation results in the trimerized triangular lattice of (b) $\kappa_{xx}$, (c) $|\kappa_{yx}|$, and (d) $\theta_H = |\kappa_{yx}|/\kappa_{xx}$ in the log-log scale. (e) The phonon spectrum of Kagome lattice with the nearest longitudinal spring constant $Q_L = 50~meV/a^2$ and the nearest transverse spring constant $Q_T = 0.3 Q_L$. The numerical computation results in the Kagome lattice of (f) $\kappa_{xx}$, (g) $|\kappa_{yx}|$, and (h) $\theta_H = |\kappa_{yx}|/\kappa_{xx}$ in the log-log scale. 
    The black double dotted lines indicate the sign change of $\kappa_{yx}$ from negative to positive at $T\approx 35~K$. Here, we assume $B_e \sim 5~meV$, and $\tau \sim 10^{-12}~s$. The blue lines are the computed results and the orange dotted lines are the approximated temperature dependence in the linear dispersion of phonons. }
    \label{fig:7}
\end{figure*}

Now, we explain how to acquire the scattering rate. We denote the original Hamiltonian $\hat H_0$ (the phonon) and the perturbative Hamiltonian $\hat V = H_R$ in Eq.~\ref{eq:15}. The Fermi golden rule approximates the scattering rate as
\alg{
\w{l'l} = 2\pi \al{|T_{l'l}|^2}_{d} \dt(\ep_{l'}-\ep_l),
}
where $T_{l'l} = \bra{l'}\hat V\ket{\psi_l}$ is the T-matrix, and $\al{}_{d}$ is the averaged value for random spin configurations.
Here, $\ket{l'}$ is the incident wavefunction which is an eigenfunction of $\hat H_0$ while $\ket{\psi_l}$ is the scattered wavefunction.
$\psi_l$ satisfies Lippmann-Schwinger equation.
\[
\ket{\psi_l} = \ket{l} + \hat G_0^l \hat V\ket{\psi_l}.
\]
where $\hat G_0^l = (\ep_l - \hat H_0 + i\eta)^{-1}$ is the Green function. The Born approximation gives
\[
\ket{\psi_l} \approx \ket{l} + \hat G_0^l \hat V\ket{l}.
\]
Thus, the T-matrix elements are
\[
T_{l'l} \approx \bra{l'}\hat V \ket{l} + \bra{l'}\hat V\hat G_0^l \hat V\ket{l}.
\]
We denote $V_{l'l} = \bra{l'}\hat V\ket{l}$. Then,
\[
T_{l'l} = V_{l'l} + \sum_{l''} \f{V_{l'l''}V_{l''l}}{\ep_l-\ep_{l''}+i\eta}.
\]
Its absolute square is approximated as
\alg{
|T_{l'l}|^2 \approx |V_{l'l}|^2 + \sum_{l''}\f{V_{ll'}V_{l'l''}V_{l''l}}{\ep_{l}-\ep_{l''}+i\eta} + c.c. + O(\hat V^4).
}
The antisymmetric part is given by
\alg{
\w{ll'}^a =& 2\pi (\sum_{l''}[(\f{\al{V_{l'l}V_{ll''}V_{l''l'}}_{d}}{\ep_{l'}-\ep_{l''}+i\eta} + c.c.)\nm
-(\f{\al{V_{ll'}V_{l'l''}V_{l''l}}_{d}}{\ep_{l}-\ep_{l''}+i\eta} + c.c.)]\dt(\ep_{l}-\ep_{l'}).
}
The leading order is symmetric while the subleading order has both symmetric and antisymmetric parts. 
That is, $1/\tau \gg |\w{ll'}^a|$, since the leading order gives $1/\tau$ and the subleading order gives $\w{ll'}^a$ in Eq.~\ref{eq:18}. In our numerical computation, we find that the average of $|V_{ll'}|^2$ is 2-order larger than that of $|\w{ll'}^a|$ for $B_e=5~meV$. Accordingly, the expectation value of $\theta_H$ can be around $10^{-2}$.

In numerical computation, we replace the delta function with
\[
\dt(\ep_{l}-\ep_{l'}) \approx -\f{1}{\pi} \Im(\f{1}{\ep_{l}-\ep_{l'}+i\eta}).
\]
Thus, we get
\alg{
\w{ll'}^a =& -2 \sum_{l''}[(\f{\al{V_{l'l}V_{ll''}V_{l''l'}}_{d}}{\ep_{l'}-\ep_{l''}+i\eta} + c.c.)\nm
-(\f{\al{V_{ll'}V_{l'l''}V_{l''l}}_{d}}{\ep_{l}-\ep_{l''}+i\eta} + c.c.)]\Im (\f{1}{\ep_{l}-\ep_{l'}+i\eta}).
}
We let $\eta = 0.01$ and $B_e \approx 5~meV$, which is the emergent field strength at the polar angle of spins $\ta \approx \pi/4$ with $t \approx 0.4$ in Fig.~\ref{fig:3}(a).

The matrix elements $\hat V_{ll'}$ can be acquired as follows.
If the triangle with the SSC is in the unit cell at $\vec{R}_i$, the perturbative Hamiltonian is written in Eq.~\ref{eq:15}.
\alg{
V_{ll'} = \int d^2r \bra{\nu_{l}(r)}\hat V(r)\ket{\nu_{l'}(r)}.
}
Here, for $l=(n,k)$, $\ket{\nu_l(r)} = e^{ik\cdot r} \ket{\chi_l(r)}$ is the Bloch wavefunction of phonons, $\ket{\chi_l} = [u_{l},-i\ep_{l} u_{l}]^T/N_{l}$ is the eigenvector of the phonon energy band, and $u_{l}$ is the polarization vector such that $D(k) u_{l} = \ep_{l}^2 u_{l}$ with the dynamical matrix $D(k)$. 
The rest of algebra is straightforward:
\alg{
V_{ll'} =& - \f{B_e}{2M} \sum_{ab} \Lambda_{ab}  (L_{ba}^z)_{ll'} e^{i(\vec{k'}-\vec{k})\cdot \vec{d}_a}.
}
Here, we call $(L_{ba}^z)_{ll'} = \bra{\chi_l}L_{ba}^z\ket{\chi_{l'}}$ the generalized angular momentum. For $(u_b^x(k), u_b^y(k),p_a^x(k),$ $p_a^y(k))$ basis, the generalized angular momentum matrix is defined as~\cite{park2020phonon}
\[
(L_{ba})^z = \f{1}{2}\left(\begin{matrix}
0 & i\sigma_y \\
-i\sigma_y & 0 
\end{matrix}\right).
\]

\subsection{The thermal conductivities}

We numerically compute the thermal conductivities and thermal Hall angle, and showcase the results of trimerized triangular lattice in Figs.~\ref{fig:7}(b-d), and those of Kagome lattice in Figs.~\ref{fig:7}(f-h).
The observed facts are the followings. First, $\kappa_{xx}$ is proportional to $T^2$ at low temperatures.
Second, $\kappa_{yx}$ is proportional to $T^3$ and $\theta_H$ is proportional to $T$ at low temperatures.
Third, for the Kagome lattice only, the sign change occurs at $T\approx 35~K$. Fourth, the range of $\kappa_{xx}$ is $10^{-2} \sim 10^{0}~W K^{-1}m^{-1}$, that of $\kappa_{yx}$ is $10^{-5}\sim 10^{-1} W K^{-1}m^{-1}$, and that of $\theta_H$ is $10^{-3} \sim 10^{-1}$. 
This is comparable with experiments, where $\kappa_{xx}$ spans from $10^0 \sim 10^1~WK^{-1}m^{-1}$, $\kappa_{yx}$ spans from $10^{-4} \sim 10^{-2}~WK^{-1}m^{-1}$, and $\theta_H$ reaches $\sim 10^{-3}$~\cite{ideue2017giant,kim2024thermal}.

We can explain the temperature dependence as follows. 
At low temperatures, the linear dispersion of phonons near the $\Gamma$ point gains importance.
If $\ep_l \approx v_n k$, $\p k_x \ep_l = v_n \cos\ta$, $\p k_y \ep_l = v_n \sin\ta$, the longitudinal thermal conductivity is
\alg{
\kappa_{xx} =& \f{\tau}{T^2}\sum_n \int \f{k dk d\ta}{(2\pi)^2} v_n^2k^2 (v_n^2 \cos^2\ta)\f{ \exp(v_nk/T)}{(\exp(v_nk/T)-1)^2}
\nq \f{\tau T^2 }{4\pi}\sum_n \int dX_n~ X_n^3 \f{ \exp(X_n)}{(\exp(X_n)-1)^2}
\nq \f{\tau T^2 }{4\pi}\sum_n 6 \zeta(3) = \f{3 \tau T^2 }{\pi} \zeta(3).
}
Here, $X_n = v_nk/T$. Thus, $\kappa_{xx} \propto T^2$. 
Also, the thermal Hall conductivity is
\alg{
 \kappa_{yx} =& -\f{\tau^2}{T^2} \sum_{nn'} \int \f{dk d\ta}{(2\pi)^2} \f{dk'd\ta'}{(2\pi)^2}v_n^2k^2 \sin\ta \w{nkn'k'}^a \cos\ta'\nm
 \times\f{v_{n'}^2k'^2 \exp(v_{n'}k'/T)}{(\exp(v_{n'}k'/T)-1)^2}. \nonumber
 }
When we numerically observe that the maximum value of $|\w{nk,n'k'}^a|$, it is nearly independent of $k$ only near the $\Gamma$ point. 
So, we can assume $\w{nk,n'k'}^a \approx N_{nn'}B_e^3\sin(\ta-\ta')\dt(v_nk - v_{n'}k')$ with the coefficient $N_{nn'}$, where the sine function comes from the antisymmetric condition. Then,
 \alg{
 \kappa_{yx} \approx & -\f{\tau^2}{T^2}B_e^3 \sum_{nn'} \f{N_{nn'}}{v_{n'}} \int \f{dk d\ta}{(2\pi)^2} \f{dk'd\ta'}{(2\pi)^2}v_n^4k^4 \sin\ta  
 \nm \times \sin(\ta-\ta')\dt(k'-\f{v_nk}{v_n'}) \cos\ta'\f{\exp(v_{n}k/T)}{(\exp(v_{n}k/T)-1)^2}
 \nq -\tau^2 T^3 B_e^3 \frac{\pi^2}{60}\sum_{nn'}\f{N_{nn'}}{v_n'v_n} . \nonumber
 }
Thus,
 \alg{
 |\kappa_{yx}| \propto \tau^2 T^3.
 }
 Accordingly, the Hall angle should be
 \alg{
 \ta_H =& \f{|\kappa_{yx}|}{\kappa_{xx}} \propto \tau T.
 }
At the higher temperature, the acoustic phonons at nonlinear dispersion participate in the transport, $\kappa_{xx},\kappa_{yx}$, and $\theta_H$ are deviated from the temperature dependence.

Also, in the Kagome lattice, we numerically observe that the sign change in $\w{ll'}^a$ occurs during the increment of energy while in the trimerized triangular lattice, this is not the case. 
This causes the sign change of $\kappa_{yx}$ only in the Kagome lattice.
We believe that the sign change of $\kappa_{yx}$ is not universal and depends on the system's microscopic detail.

Physically, the skew-scattering arises as follows. [See Fig.~\ref{fig:1}.]
With the time-reversal and inversion symmetry, the left and right circularly polarized chiral phonons are degenerate in the $k$-space.
The atoms rotating in the clockwise (counterclockwise) direction form the left (right) circularly polarized chiral phonons. 
In general, left and right circularly polarized chiral phonons have opposite generalized angular momenta $L_{ba}$.
When the emergent field from the local SSC is there, the time-reversal symmetry is broken, and this induces the THE due to phonon skew-scattering. 
This is analogous to the skew-scattering induced anomalous Hall Effect by magnetic impurities~\cite{nagaosa2010anomalous}.

\section{Discussions \label{sec:6}}

We discuss the typical values estimated in our works. For the setting parameters, the double exchange coupling constant is $J_H \sim 1~eV$, the emergent field is $B_e \sim 5~meV$, the logitudinal spring constant is $Q_L \sim 50~meV/a^2$ with lattice constant $a$, and the relaxation time $\tau \sim 10^{-12}~s$. 
Accordingly, near $T \approx 60 \sim 80~K$, $\kappa_{xx} \approx 10^{-1} \sim 10^{0}~ W K^{-1} m^{-1}$ $\kappa_{yx} \approx 10^{-3}\sim 10^{-2}~ W K^{-1} m^{-1}$, and $\ta_H \approx 10^{-3} \sim 10^{-2}$ on the trimerized triangular (YMnO$_3$) and Kagome lattices. 
This gives much larger value than the thermal Hall conductivity $\kappa_{yx} \sim 10^{-6} ~W K^{-1} m^{-1}$ from the spin-phonon interaction via spin-orbit coupling~\cite{saito2019berry}.

The candidates of this phenomenon are the general Mott insulating systems that can host the SSC fluctuation. 
The typical examples of the trimerized triangle lattices are YMnO$_3$~\cite{kim2024thermal} and LuMnO$_3$~\cite{kim2019magnon}.
The examples of Kagome lattices are YCu3-Br~\cite{zheng2023unconventional} and Na$_2$Mn$_3$Cl$_8$~\cite{paddison2023multiple}. 
The cuprates~\cite{boulanger2020thermal} are also  candidates.
There could be more numerous examples that we do not mention.

So far, we have discussed that the THE can be induced by the phonon skew-scattering by the SSC. 
The electronic many-body wavefunction is deformed by the SSC, giving rise to the emergent Raman interaction analogous to the Aharanov-Bohm Effect. 
The emergent Raman interaction skew-scatter the chiral phonons with the opposite angular momenta, inducing the THE that can be measured in experiments. 
By revealing the brand-new phonon skew-scattering mechanism, we believe that our work expands the window of noncoplanar spin structures and fluctuations with their correlations with the phonons.

\begin{acknowledgments}

We send our sincere acknowledgment to Hiroki Isobe, Yingming Xie, and Wataru Koshibae for fruitful discussions. T.O. and N.N. were supported by JSPS KAKENHI Grant Numbers 24H00197 and 24H02231, and the RIKEN TRIP initiative.

\end{acknowledgments}

\bibliography{reference.bib}


\end{document}